# Which is conducive to high responsivity in a hybrid graphene-quantum dot transistor: Bulk- or layer-heterojunction

Yating Zhang[1, 2 *], Xiaoxian Song[1, 2], Ran Wang[1, 2], Haiyan Wang[1, 2], Yongli Che[1, 2], Xin Ding[1, 2], and Jianquan Yao[1, 2]

[1] Institute of Laser & Opto-Electronics, College of Precision Instruments and Opto-electronics Engineering, Tianjin University, Tianjin 300072, China

[2] Key Laboratory of Opto-electronics Information Technology (Tianjin University), Ministry of Education, Tianjin 300072, China

**Corresponding author**: yating@tju.edu.cn

**Abstract.** Heterojunction, a photoelectric conversion center, plays a critical role in photo detection. A compare study is performed on field effect phototransistors (FEpT) with two typical heterojunctions based on graphene and PbSe quantum dots (QDs) hybrid, including layer-heterojunction (LH) and bulk-heterojunction (BH). LH-FEpT exhibits a higher mobility ($\mu_n$) 677 $cm^2V^{-1}s^{-1}$, whereas $\mu_n$ of BH-FEpT is as low as 314 $cm^2V^{-1}s^{-1}$, however the higher responsivity ($10^6$ A/W) is achieved by the former. Because R is proportional to the product of mobility and transfer rate, and high transfer rate of LH-FEpT overcompensates the shortage of low mobility. Although large area of heterojunction and high mobility benefit high density of photo-induced carriers, lack of transport mechanism becomes to the main constrain factor in BH-FEpT. Therefore, LH-FEpT is a better candidate of near infrared (NIR) photo detectors.

Heterojunction, a photoelectric conversion center, plays a critical role in photo detection. Two typical heterojunctions are layer-heterojunction (LH) and bulk-heterojunction (BH), which have both advantages for sensing light. However, which type achieves higher responsivity is still a problem. Here, a compare study is performed on these two typical heterojunctions based on graphene and PbSe quantum dots (QDs) hybrid. By investigation of the output, and transfer characteristics, the responsivities and response time, we conclude that LH-FEpT achieves higher response rate and responsivity, therefore becomes a better candidate of NIR photo detectors.

1. **Introduction**

Solution-processed FEpTs based on graphene, inorganic quantum dots (QDs) and their hybrid have achieved in ultrahigh performances [1-6], and are attracting more and more attentions, due to great potential applications in NIR detectors with low cost, low energy-consumption, flexibility, easy fabrication and easy integration[7-10]. As a center of exciton (electron-hole pair) separation or carrier transfer, heterojunction plays a critical role in photo detection [11, 12]. In general, there are two typical heterojunctions for FEpT channel structures: layer-heterojunction (LH) and bulk-heterojunction (BH). Based on these structures, photo responses have been reported. Xu's group reported BH-FEpTs based on grphene and lead sulfide quantum dots (PbS QDs) hybrid[13]. Under gate voltage of -10 V and bias voltage of 100 μV, photoresistance reached to 20 Ω [13]. Konstantato's group and Yan's group both reported LH-FEpT based on graphene and PbS colloidal quantum dots hybrid with ultrahigh responsivity of $10^7$ A/W [14, 15].

The responsivity $R_{ph}$ of a thin film photoconductor is expressed as $R_{ph} = en\mu EW/P$, where e is electronic charge, n is the density of photo-induced carriers per unit area, μ is the carrier mobility, E is the applied electric field, P is incident optical power, and W is the width of the device[15]. Based on that, BH has larger interface where exciton separates (high n), but irregular shape of channels where carriers transport inefficiently[16]; LH has regular shape of channel, but small interfacial area (low n) [14, 15]. However, considering above factors, which type achieves higher responsivity is still a problem.

Here, two typical FEpTs are fabricated based on graphene and PbSe QDs hybrid. By a compare study, we investigated the output, and transfer characteristics, the responsivities and response time of them. We conclude that LH-FEpT achieves higher response rate and responsivity, therefore becomes a better candidate of NIR photo detectors.

2. **Experiments**

2.1 Materials

Graphene powder used in BH-FEpT was purchased from The Sixth Element Inc, and single layer graphene on Si n $^+$/SiO2 used in LH-FEpT was purchased from Hefei Weijing Material Technology Co., Ltd. PbSe QDs was synthesized through a wet chemical method [17]. PbSe QDs was characterized by optical spectra and transmission electron microscope (TEM), as the insert of figure 1 (a) with a scale bar of 20 nm. The absorption and photoluminescence (PL) spectra were performed on PbSe QDs toluene solution by Zolix Omni-λ300 spectrometer. The absorption peak locates at 1703 nm, while PL peak is at 1730 nm excited by a cw laser in wavelength of 532 nm. According to a four-band-envelope-function formulism, the average diameter of PbSe QDs is 6.5 nm[18]. High resolution transmission electron microscopy (HRTEM) was performance by transmission electron microscope from FEI Co., Tecnai G2 F20, and 200 kV. Based on TEM image, the average size of the PbSe QDs is attributed to 6.5 nm consistence with the size deduced from absorption spectrum of figure 1 (a).

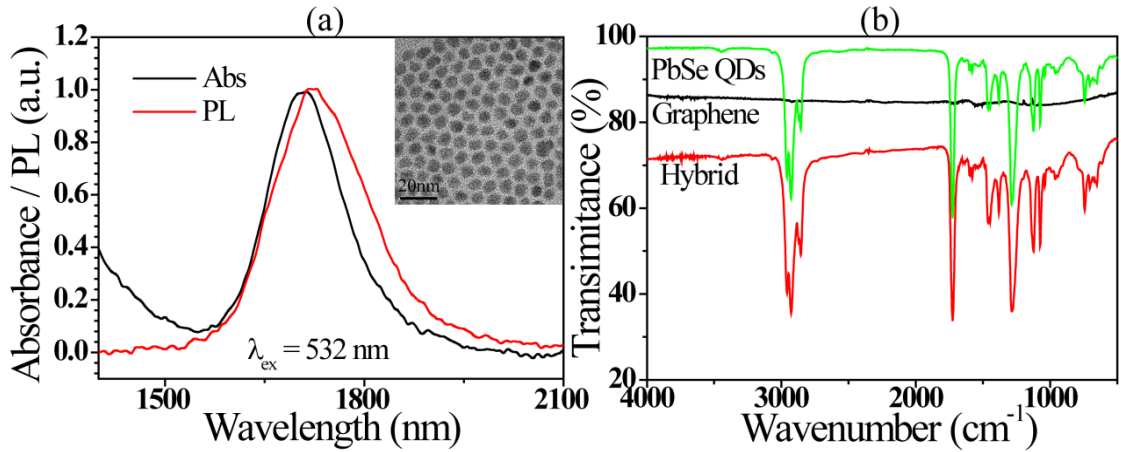

**Figure 1** (a) the absorption and photoluminescence (PL) spectra of PbSe QDs, the insert is HRTEM image of PbSe QDs with a scale bar of 20 nm. (b) FTIR spectra of PbSe QDs, Graphene, and the hybrid.

The hybrid solution was prepared by combining one volume of graphene toluene solution (~1 mg/ml) and three volumes of PbSe QDs (10 mg/ml) toluene solution. The transmission peaks of PbSe QDs, MEH-PPV, and their hybrid have been characterized through fourier transform infrared (FTIR) spectra (by FTIR-650-spectrometer from Tianjin Gangdong sci.&tech. development Co., Ltd.), as figure 1 (b). For PbSe QDs, FTIR spectra exhibits the properties of ligand, which is attributed to dioctyl phthalate (DOP), and no peaks was from graphene in such region. We believe that DOP is a by-product of PbSe QDs reaction, which is capping the surface of QDs when prepared. For the hybrid, the characteristic peaks are the superposition of graphene and PbSe QDs.

2.2 Fabrication of FEpTs

LH- and BH- FEpTs based on graphene-PbSe QD hybrid were fabricated experimentally. The schematic diagrams are displayed in figure 2 (a) and (b), respectively. Substrates are Si $n^+$/$SiO_2$. The thickness of $SiO_2$ is 300 nm.

LH-FEpT was fabricated by following processes. A single layer graphene was transferred to the surface of substrate (the growth and transfer of single layer graphene were done by Hefei Weijing Material Technology Co., Ltd). Then Au source and drain in thickness of 200 nm were thermally evaporated over a shadowed mask. Three layers of PbSe DQs were deposited layer by layer. Each layer was prepared as follows: 2% (Vol.) ethanedithiol (EDT) in acetonitrile solution was first prepared for ligand exchange. PbSe QDs toluene solution was prepared at 10 mg/mL in toluene. A drop of PbSe layer was first deposited on the spinning substrate with speed of 2000 rpm and left for 15 seconds to dry. 3 drops of 2% EDT solution were then deposited on the rotating substrate and followed by 2 drops of acetonitrile and 2 drops of toluene.

The BH-FEpT was fabricated by Au electrodes thermal evaporation directly on the substrate over the same shadowed mask. The thickness was 200 nm. Three layers of graphene-QD hybrid were spin casted from the hybrid solution, as figure 2 (b). Each layer was deposited by a similar process to LH-FEpT layer preparation, except for replacing PbSe solution by hybrid solution. Each device was dried in vacuum overnight.

The SEM images of cross sections of LH-FEpT and BH-FEpT are given by figure 2 (c) and (d), respectively. The structure of each device is very clearly visible. The bottom layer is Si $n^+$, on which is $SiO_2$ layer. The thickness is attributed to 300 nm. The top layer is heterojunction layer. The thickness of PbSe QDs layer in LH-FEpT is 75 nm, and that of BH layer is 84 nm. Clearly, delamination is absent or present in BH and LH-FEpT, respectively.

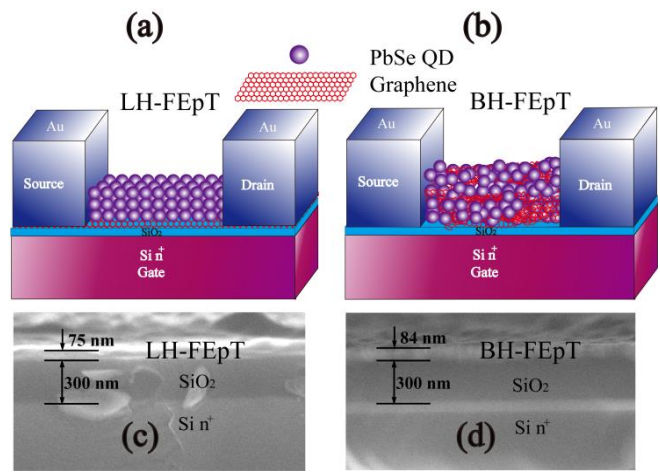

**Figure 2** Schematic diagram (a) and the cross-sectional SEM image (c) of LH-FEpT: a singal layer of graphene and three layers of PbSe QDs. Schematic

diagram (b) and the cross-sectional SEM image of BH-FEpT (d): three layers of the graphene - PbSe QDs hybrid.

2.3 Electrical measurement

In electrical measurement, bias voltage ($V_{SD}$) was applied over source (ground connection) and drain electrodes by Keithley 2400, channel current flowing into drain denoted by $I_D$ was also detected by Keithley 2400. Gate voltage ($V_G$) was applied on gate electrode by HP6030A to ground connection.

3. **Results and discussion**

3.1 Electric properties of FEpTs

Electric measurements were performed at ambient atmosphere under room temperature in the dark. Figure 3 (a) and (b) give the ($I_D$ - $V_{SD}$) output characteristics of LH- and BH-FEpTs, respectively, under some certain bias voltages of 0 V, ±2 V, ±4 V, ±6 V, and ±8 V. Ambipolar characteristics of both devices are obvious saturation behavior in both hole and electron depletion regimes in graphene[19]. Drain current ($I_D$) for both p- and n-channel operations shows a nearly linear increase in the $V_{DS}$ range of 0 to ±5 V for both devices. In such region the drain current is expressed as $I_D = n_G en\mu V_{SD}W/L$ [19-22], wherer $n_G$ is number of graphene sheets forming the channel, e is the electronic charge, n is carrier density per area in graphene sheet(s), $V_{SD}$ is bias voltage, μ is carrier mobility in graphene sheet(s), W and L are width and length of the channel. A higher n is obtained by applied a higher $V_G$, because more carriers are driven from QDs to graphene sheet(s) [15]. Therefore, the higher of $V_G$ applied the larger of channel current is. The saturation current indicates the transfer rate of corresponding charge carriers under $V_G$. For BH-FEpT, the same level of saturation current in n- and p- channel operations represents that electrons and holes have balance transfer rate; whereas for LH-FEpT, hole transfer rate is higher than that of the electrons by deducing from higher level of saturation current of holes.

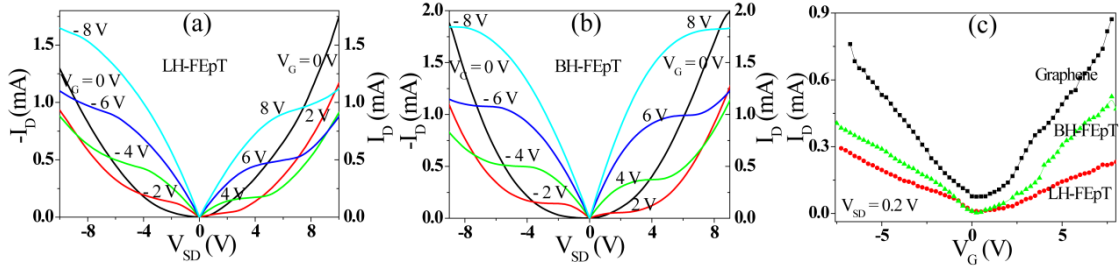

**Figure 3** Output characteristics ($I_D - V_{SD}$) of LH-FEpT (a) and BH-FEpT (b) without light illumination, under $V_G$ of 0 V, ±2 V, ±4 V, ±6 V, and ±8 V. (c)Transfer characteristics ($I_D - V_G$), $V_{SD} = 0.2$ V of graphene FET, BH-FEpT and BH-FEpT.

The mobility of electron and hole as another important parameter in FEpTs can be obtained by transfer characteristics ($I_D - V_G$), as figure 3 (c). The symmetry of transfer curve of pristine graphene FET indicates equal mobilities of hole and electron which are determined by the slope of transfer curve[15]. The minimum point corresponds to the lowest carrier density in graphene, or termed as Dirac point. Modified by QDs, the Dirac point shift is very small for both devices, which is different from the result of Fan's group[15], due to the barrier formed by capping ligands preventing carriers from transferring. When gate voltage is applied, carrier transfer occurs, then the polarity and carrier density in graphene sheet(s) can be tuned [23, 24]. According to the slopes in figure 3 (c), mobilities (M) of electron and hole are attributed to 314, 367 $cm^2$/Vs in BH-FEpT, and 677, 527 $cm^2$/Vs in LH-FEpT, respectively. Please note that M is mobility of carrier in channel, instead of mobility μ in graphene sheet(s), and $M = n_G\mu$. Where $n_G$ can be regarded as sub-circuit number in a shunt circuit, it reduce to 1 if single layer graphene forming the channel as LH-FEpT. Thus, high mobility of carriers in LH-FEpT is the consequences of multi-pathways formed by multi-graphene sheets in channel.

3.2 Photo responsivities of FEpTs

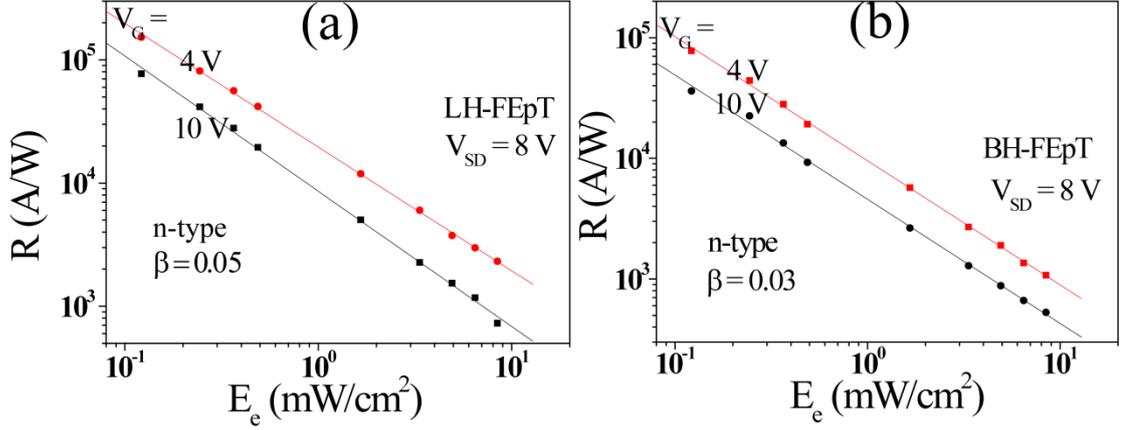

**Figure 4** Responsvities of LH-FepT and BH-FepT in n-channel operating regime (a) and (c), respectively.

Sensing mechanism consist of three links. Firstly, photo induced carriers generate in QDs. Then, some of them transfer to graphene driven by gate voltage, and transport in graphene under bias voltage. Meanwhile, carries in opposite charge remain in QDs can be viewed as an additional gate voltage applied on graphene, which termed as *light-gate effect*. Due to that effect, horizontal shifts $\Delta V_G$ could be used to calibrate light irradiance ($E_e$), as [25]

$$\Delta V_G = \alpha E_e^{\beta} \tag{1}$$

where α and β are constant. According to equation $I_D = WC_{ox}V_{SD}\mu(V_G - V_T)/L$ [20, 21], the increments of channel current $\Delta I_{SD}$, resulting from the light illumination, could be expressed as a function of the gate shift $\Delta V_G$ deduced from equation (1):

$$\Delta I_D = \frac{W}{L} C_{ox} \mu V_{SD} \Delta V_G \tag{2}$$

For a specific device, $WC_{ox}\mu/L$ is constant, and $\Delta I_{SD}$ is directly proportion to $V_{SD}\Delta V_G$.

Responsivity (R) is also calculated by equation

$$R = \frac{I_{ill} - I_{Dark}}{P} = \frac{\Delta I_D}{P} \tag{3}$$

where $I_{ill}$ and $I_{Dark}$ are channel current under light illumination and in dark, respectively, and $P = AE_e$ is incident optical power, where A is the illumination area. Substituting equation (1) (2) into (3), obtain expression

$$R = \frac{C_{ox}V_{SD}\mu}{L^2 E_e}\Delta V_G = \frac{\alpha C_{ox}V_{SD}\mu}{L^2} E_e^{\beta-1} \qquad (4a)$$

$$\lg(R) = \lg\left(\frac{\alpha C_{ox}V_{SD}\mu}{L^2}\right) + (\beta-1)\lg(E_e) \qquad (4b)$$

In a double-logarithmic axis plot, lg(R) keeps a good linear relation with (β-1)lg($E_e$) for both types of FEpTs, consistence with results of other groups [14, 15]. Both of FEpTs exhibit photo responses as shown in figure 4, under the specific gate voltage (10 V, and 4 V, ) and bias voltage (8 V). Ambipolar photo response indicates that both types of operation channels can convert light-gate voltage to light-drain current. The slope of lg(R) versus lg($E_e$) is attributed to a consistence (β-1). Fitting with equation (4b), we got, $\beta_n$ = 0.05 for LH-FEpT, and $\beta_n$ = 0.03 for BH-FEpT. The highest β and responsivity presents in LH-FEpT when operating in n-channel corresponding the lowest mobility of 314 cm$^2$/Vs.

3.3 Light response time of FEpTs

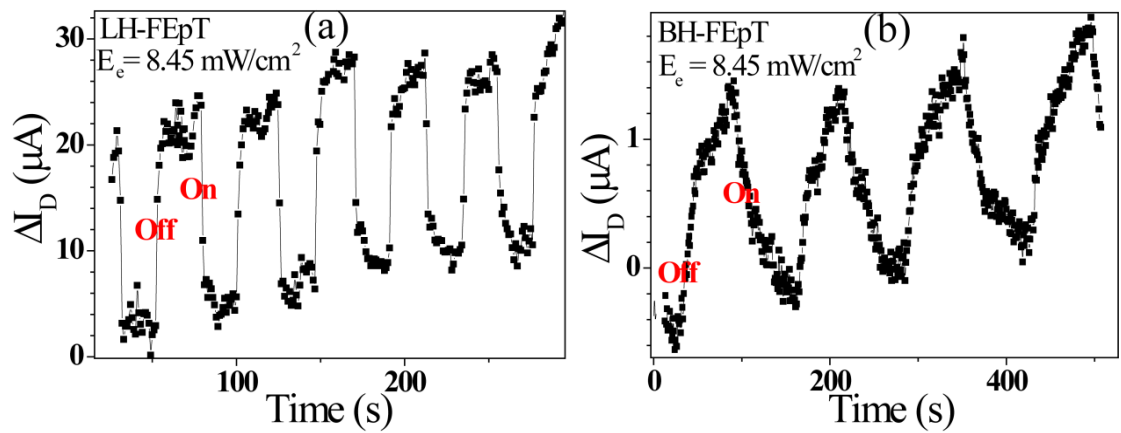

**Figure 5** Photo current responses of LH-FEpT (a), and BH-FEpT (b) to On/Off light illumination for various cycles. $V_G$ = -5, $V_{SD}$ = 1 V; Wavelength: 808 nm; Irradiation: 8.45 mWcm$^{-2}$; On / Off time: 20 s (a), and 75 s (b)

Figure 5 gives the transient photo current response of two FEpTs to On / Off illumination for various cycles, where light irradiation is 8.45 mW/cm$^2$, wavelength is 808 nm, $V_G$ = 5, $V_{SD}$ = 1 V, and On / Off time is 20 s and 75 s for LH-FEpT and BH-FEpT, respectively. The channel current decrease with illumination time, and can be fitted with an exponential equation: $\Delta I_D = \Delta I_1 \exp(-t/\tau_1) + \Delta I_2 \exp(-t/\tau_2)$ with two relaxation time $\tau_1$ and $\tau_2$. The time constant $\tau_1$ and $\tau_2$ are 0.7 s and 2.5 s for LH-FEpT, 12 s and 49 s for BH-FEpT, respectively. The shorter relaxation time $\tau_1$ indicates transfer time of holes from QDs to graphene sheet(s), whereas the longer response time $\tau_2$ corresponds to carrier transport time in QD layer. Similarly, the channel current increases when the light is switched off. When fitting by an exponential equation with two relaxation time $\tau_3$ and $\tau_4$: $\Delta I_D = \Delta I_1 (1-\exp(-t/\tau_3)) + \Delta I_2 (1-\exp(-t/\tau_4))$, the time constant $\tau_3$ and $\tau_4$ are 1 s and 3 s for LH-FEpT, 15 s and 58 s for BH-FEpT, which are slightly longer than $\tau_1$ and $\tau_2$, respectively. Considered $\tau_3$ represents transfer time of electrons from QDs to graphene sheets, $\tau_1 < \tau_3$, indicates QDs are net negative charged when exposed to light. Though dark current shifts, photo current of both devices is reproducible.

Obviously, all time constant $\tau_1$ - $\tau_4$ of LH-FEpT is much shorter than that of BH-FEpT. The most import time constant is $\tau_1$ whose inverse represents the transfer rate from QDs to graphene. According to fitting results $\tau_1$ are 0.7 s and 12 s, the transfer rate of LH-FEpT and BH-FEpT is 1.43 and 0.083 for LH-FEpT and BH-FEpT, respectively. In other words, the ratio of transfer rate of LH-FEpT to BH-FEpT is as high as 17.

3.4 Sensing mechanism discussion

According to sensing mechanism, those photon-generated carriers transfer from PbSe QDs to graphene form the $\Delta V_G$, representing in unit time photo-induced additional gate voltage, expressed as [20, 21]

$$\Delta V_G = \int_V \frac{er_{tr} n_{ph}}{C_{ox} WL} dV \qquad (5)$$

where $r_{tr}$ is net transfer rate, $n_{ph}$ is photo-induced carrier density per unit volume in channel, and integration is performed on channel volume. However, $\Delta V_G$ cannot be measured directly. Then $\Delta I_D$ is used as an observation variable associated with $\Delta V_G$ by equation (2). Substituting equation (2) (5) into equation (3), we obtain R as

$$R = \frac{\Delta I_D}{P} = \frac{eV_{SD}}{L^2 P} \mu \int_V r_{tr} n_{ph} dV \qquad (6)$$

Setting $\int_V r_{tr} n_{ph} dV = \bar{r}_{tr} N_{ph}$, where $\bar{r}_{tr}$ is the average net transfer rate, $N_{ph}$ is the total number of photo-induced charge carriers. Then, a more simple expression of R is got,

$$R = \frac{\Delta I_D}{P} = \left(\frac{eV_{SD}}{L^2 P} N_{ph}\right) \mu \bar{r}_{tr} \qquad (7)$$

From equation (7), it is obvious that responsivity is proportional to the product of mobility and average net transfer rate in this kind of hybrid FEpTs. In our experiment, $\mu_n$ of LH-FEpT is nearly a half of value in BH-FEpT, meanwhile $\bar{r}_{tr}$ of LH-FEpT is about 17 times that of BH-FEpT. As a consequence, R of LH-FEpT is deduced to much higher than R of BH-FEpT, consistence with the results shown in figure 4.

4. **Conclusions and summary**

In summary, we investigated two typical FEpTs, based on graphene and PbSe QDs hybrid, including LH-FEpT and BH-FEpT. Due to multi-pathways formed by multi-graphene sheets in bulk-heterojunction channel, BH-FEpT shows the highest mobility $\mu_n = 677$ cm$^2$V$^{-1}$s$^{-1}$, whereas $\mu_n$ in LH-FEpT is as low as 314 $cm^2V^{-1}s^{-1}$. The higher responsivity is obtained from LH-FEpT, instead of BH-FEpT. Because R is proportional to the product of mobility and transfer rate, and high transfer rate of LH-FEpT overcompensates the shortage of low mobility. Although large area of BH benefits high density of photo-induced carriers, lack of transport mechanism becomes to the main constrain factor. Therefore, LH-FEpT is a good candidate of NIR photo detector with low cost, flexibility, easy fabrication and easy integration.

This work is supported by the National Natural Science Foundation of China (Grant No. 61271066) and the Foundation of Independent Innovation of Tianjin University (Grant No. 60302070).

**References**

1. Yang, Y., W. Rodriguez-Cordoba, and T. Lian, *Multiple Exciton Generation and Dissociation in PbS Quantum Dot-Electron Acceptor Complexes.* Nano Letters, 2012. **12**(8): p. 4235-4241.

2. McDonald, S.A., et al., *Solution-processed PbS quantum dot infrared photodetectors and photovoltaics.* Nat Mater, 2005. **4**(2): p. 138-142.

3. Zhang, S., et al., *Enhanced infrared photovoltaic efficiency in PbS nanocrystal/semiconducting polymer composites: 600-fold increase in maximum power output via control of the ligand barrier.* Applied Physics Letters, 2005. **87**(23): p. -.

4. Konstantatos, G., et al., *Ultrasensitive solution-cast quantum dot photodetectors.* Nature, 2006. **442**(7099): p. 180-183.

5. Rauch, T., et al., *Near-infrared imaging with quantum-dot-sensitized organic photodiodes.* Nature Photonics, 2009. **3**(6): p. 332-336.

6. Zhang, W., et al., *Ultrahigh-Gain Photodetectors Based on Atomically Thin Graphene-MoS2 Heterostructures.* Sci. Rep., 2014. **4**.

7. Hetsch, F., et al., *Quantum dot field effect transistors.* Materials Today, 2013. **16**(9): p. 312-325.

8. Yu, H., Z. Bao, and J.H. Oh, *High-Performance Phototransistors Based on Single-Crystalline n-Channel Organic Nanowires and Photogenerated Charge-Carrier Behaviors.* Advanced Functional Materials, 2013. **23**(5): p. 629-639.

9. Feng, W., et al., *A layer-nanostructured assembly of PbS quantum dot/multiwalled carbon nanotube for a high-performance photoswitch.* Scientific Reports, 2014. **4**.

10. Kramer, I.J. and E.H. Sargent, *The Architecture of Colloidal Quantum Dot Solar Cells: Materials to Devices.* Chemical Reviews, 2014. **114**(1): p. 863-882.

11. Zhou, Y., M. Eck, and M. Kruger, *Bulk-heterojunction hybrid solar cells based on colloidal nanocrystals and conjugated polymers.* Energy & Environmental Science, 2010. **3**(12): p. 1851-1864.

12. Liao, H.-C., et al., *Nanoparticle-Tuned Self-Organization of a Bulk Heterojunction Hybrid Solar Cell with Enhanced Performance.* Acs Nano, 2012. **6**(2): p. 1657-1666.

13. Huang, Y.Q., et al., *Photoelectrical response of hybrid graphene-PbS quantum dot devices.* Applied Physics Letters, 2013. **103**(14).

14. Konstantatos, G., et al., *Hybrid graphene-quantum dot phototransistors with ultrahigh gain.* Nat Nano, 2012. **7**(6): p. 363-368.

15. Sun, Z., et al., *Infrared Photodetectors Based on CVD-Grown Graphene and PbS Quantum Dots with Ultrahigh Responsivity.* Advanced Materials, 2012. **24**(43): p. 5878-5883.

16. Mingqing, W. and W. Xiaogong, *P3HT/TiO2 Bulk Heterojunction Solar Cell Sensitized by Copper Phthalocyanine*, in *Proceedings of ISES World Congress 2007 (Vol. I – Vol. V)*, D.Y. Goswami and Y. Zhao, Editors. 2009, Springer Berlin Heidelberg. p. 1303-1307.

17. Kigel, A., et al., *PbSe/PbSexS1−x core-alloyed shell nanocrystals.* Materials Science and Engineering: C, 2005. **25**(5–8): p. 604-608.

18. Kang, I. and F.W. Wise, *Electronic structure and optical properties of PbS and PbSe quantum dots.* Journal of the Optical Society of America B, 1997. **14**(7): p. 1632-1646.

19. Bisri, S.Z., et al., *High Performance Ambipolar Field-Effect Transistor of Random Network Carbon Nanotubes.* Advanced Materials, 2012. **24**(46): p. 6147-6152.

20. Sze, S.M. and K.K. Ng, eds. *Physics of Semiconductor Devices*. third ed. Vol. Chapter 6. 2007, A JOHN WILEY & SONS, JNC., PUBLICATION: Canada. 293.

21. Neamen, D.A., *Semiconductor Physics and Devices: Basic Principles*. Third ed, ed. D.A. Neamen. Vol. Chapter 11 and Chapter 12. 2003: Elizabeth A. Jones.

22. Kim, C.O., et al., *High photoresponsivity in an all-graphene p–n vertical junction photodetector.* Nat Commun, 2014. **5**.

23. Rath, A.K., et al., *Solution-Processed Heterojunction Solar Cells Based on p-type PbS Quantum Dots and n-type Bi2S3 Nanocrystals.* Advanced Materials, 2011. **23**(32): p. 3712-+.

24. Zarghami, M.H., et al., *p-Type PbSe and PbS Quantum Dot Solids Prepared with Short-Chain Acids and Diacids.* Acs Nano, 2010. **4**(4): p. 2475-2485.

25. Britnell, L., et al., *Strong Light-Matter Interactions in Heterostructures of Atomically Thin Films.* Science, 2013. **340**(6138): p. 1311-1314.